\documentclass{PoSpreprint}

\usepackage{bbold}
\usepackage{slashed}
\usepackage{amsmath}
\usepackage{amssymb}

\newcommand{\be}{\begin{equation}}
\newcommand{\ee}{\end{equation}}
\newcommand{\eps}{\varepsilon}
\newcommand{\MPl}{M_{\rm Pl}}
\newcommand{\gCS}{g$\mathcal C\mathcal S\ $}

\title{Gravitational Anomalies in string-inspired Cosmologies: from Inflation to  Axion Dark Matter?}

\ShortTitle{Anomalies and Axions}

\author{\speaker{Nikolaos E. Mavromatos}\thanks{Work supported in part by the STFC (UK) Research Council through the grants ST/P000258/1 and ST/T000759/1, 
and  by the COST Association Action CA18108 ``Quantum Gravity Phenomenology in the Multimessenger Approach (QG-MM)''. NEM also acknowledges a scientific associateship (``Doctor Vinculado'') at IFIC-CSIC-Valencia University, Valencia, Spain.}\\
        Theoretical Particle Physics and Cosmology Group, Department of  Physics, King's College London, London WC2R 2LS, UK\\
        E-mail: \email{Nikolaos.Mavromatos@kcl.ac.uk}}

\abstract{In this talk, I review briefly a scenario for the evolution of a string-inspired cosmological model, in which condensates of primordial gravitational waves (GW), formed at the very early eras after the Big Bang, are considered responsible for inducing inflation and then a smooth exit to a radiation dominated epoch. Primordial axion fields, that exist in the fundamental massless gravitational (bosonic) string multiplet, couple to the non-trivial GW-induced anomalies. As a result of this coupling, there exist axion background configurations which violate (spontaneously) Lorentz symmetry, and 
remain undiluted at the end of inflation. In models with heavy sterile right-handed neutrinos (RHN), such backgrounds are linked to novel (Lorentz and CPT Violating) mechanisms for the generation of matter-antimatter asymmetry in the Cosmos, via the asymmetric decays of the RHN to standard model particles and antiparticles. During the QCD epoch, the axions develop an instanton-induced mass and can, thus, play the r\^ole of Dark Matter (DM). The energy density of such a Universe, throughout its evolution, has the form of that of a ``running vacuum model'', that is, it can be expanded in power series of even powers of the Hubble parameter $H(t)$. The coefficients of those terms, though, are different for the various cosmological epochs. For the phenomenology of our model, which is consistent with the current cosmological data, and could also help in alleviating (some of) the tensions, it suffices to consider up to and including quartic powers of $H(t)$. In the early Universe phase, it is the $H^4(t)$ term, induced by the GW condensate of the gravitational anomaly, that drives inflation without the need for external inflaton fields.}

\FullConference{Corfu Summer Institute 2019 "School and Workshops on Elementary Particle Physics and Gravity" (CORFU2019)\\
		31 August - 25 September 2019\\
		Corfù, Greece}

\begin{document}

\section{Introduction and Motivation \label{sec:intro}}

Although the cosmic concordance framework for cosmology, or $\Lambda$CDM, involving a $\sim 69\%$ current-era dark energy of positive cosmological-constant (de Sitter $\Lambda > 0$) type, a $\sim 27\%$ Cold Dark Matter (CDM) component, and $\sim 4\%$ ordinary (mostly baryonic) matter, appears to describe the large scale Universe satisfactorily, within the current accuracy~\cite{planck}, nonetheless tensions started to emerge in the data. One type of tension ($H_0$ tension) is associated with more precise measurements of the Hubble parameter in our local celestial neighbourhood (cepheids {\it etc.})~\cite{tensionH0}, whose value is found to be different from that measured by the Planck collaboration~\cite{planck}, based on Cosmic Microwave Background (CMB) studies;  there is also the $\sigma_8$ tension~\cite{s8}, associated with large-scale-formation data, and in particular with the measurement of the root-mean-squared matter fluctuation at distance scales 8h$^{-1}$~Mpc (with h the reduced Hubble parameter)~\cite{planck}. The latter is found to be less than the one predicted within the $\Lambda$CDM framework, pointing to the fact that, if such tensions survive future, more accurate, measurements,  the observed large scale structure cannot be described well by the standard concordance model and new physics might be needed for an explanation. The $\Lambda$CDM model is also found to be in tension with observations at smaller (galactic) scales, in the sense that, numerical simulations on the growth of the Universe structures at galactic scales based on it fail to describe observations, a problem known as ``small-scale cosmology crisis''~\cite{smallscale}. If such small-scale discrepancies persist  in future observations, and survive more mundane, astrophysical, explanations,  then 
they may also point towards a departure from the conventional CDM model of Dark Matter (DM), by the inclusion, for instance, of 
DM self-interactions~\cite{idm}.

That the cosmic-concordance ($\Lambda$CDM) framework cannot be the end of the story in the description of the Universe should have been expected from a theoretical point of view. There is no microscopic framework underlying $\Lambda$CDM, and, most likely, an understanding of the nature of the  dark energy will come once we start having an idea about the theory behind the quantisation of the gravitational interaction, the elusive {\it quantum gravity}. To date, the most successful attempts towards the construction of a consistent theory of quantum gravity can be found within the framework of (super)string theory and its brane extensions~\cite{strings}. It goes without saying that the above statement should not be interpreted as implying that string/brane theories are free from ambiguities or shortcomings. On the contrary, despite great effort and progress made over the past four decades, there are still many unresolved issues or unsatisfactory features of strings, such as, for instance, the lack of uniqueness of the vacuum state (String Landscape), which results in anthropic principle arguments, should one attempt to explain our observable world in this framework. No one knows whether this problem can be avoided by some new, yet undiscovered, principle that selects the true (non pertrubative) vacuum of the string/brane theories. Nevertheless, within such a framework, one has made significant progress towards: a microscopic understanding of the unification of the three known forces of Nature with gravity, the nature of 
quantum states of certain objects that belong to the realm of quantum gravity, such as 
black holes, wormholes {\it etc}., as well as the construction of natural extensions of General Relativity (GR) and Cosmology, given that the low-energy limit of string/brane theories provides extensions of GR, involving higher powers of the Riemann curvature terms and, in general, higher derivative expansions, and modifications coming from the extra (bulk) dimensions of space, which such theories live on. Such GR extensions incorporate naturally (time-dependent) dark energy, and in general distinct deviations from $\Lambda$CDM with, in principle, testable predictions. From the particle physics point of view, string theory provides naturally a plethora of DM candidates, such as supersymmetric partners, axions {\it etc.} Several consistent models also entail non thermal DM.  

In the above context, a rather simple string-inspired cosmological model has been suggested in ref.~\cite{bms}. The novel character of this model, as compared with other string cosmologies, is that it places special emphasis on the r\^ole of primordial gravitational anomalies in inducing inflation itself, and subsequent smooth exit from it into a radiation phase of the Universe and generation of matter-antimatter asymmetry in the Cosmos. A key ingredient in the model is the induction at a very early phase after the Big Bang of a gravitational-anomaly condensate due to primordial gravitational-wave (GW) fluctuations of the space-time metric. It is such a condensate that induces inflation. The presence of the condensate also implies the existence of a Lorentz-violating (LV) background configuration for primordial axion fields that exist in string theory models as part of the massless (bosonic) gravitational multiplet. Such backgrounds remain undiluted at the end of inflation and  are thus present during the radiation era~\cite{bms}. In models involving heavy right-handed neutrinos (RHN), such LV axion backgrounds can lead to the generation of matter-antimatter asymmetry, via baryogenesis through leptogenesis induced by the decays of RHN into Standard Model (SM) particles and antiparticles in such backgrounds~\cite{boss}. Moreover, during the QCD epoch of the Universe, the axions can develop a potential induced by instantons, and thus a mass, thereby providing a source for DM~\cite{bms}. 

One of the most important features of the model of \cite{bms} is that the corresponding vacuum energy density acquires a form that resembles that of the so-called ``running vacuum model'' (RVM)~\cite{sola}, which is known to provide an effective theory framework  for a smooth cosmological evolution from inflation to the current era~\cite{history}. According to the RVM, 
the energy density of the cosmic {\it vacuum} fluid obeys a 
renormalization group equation (RGE) as a function of the Hubble rate $H$~\cite{sola}:
\begin{equation}\label{rge}
\frac{d\, \rho^\Lambda_{\rm RVM} (t)}{d\, {\rm ln}H^2 } = \frac{1}{(4\pi)^2}
\sum_i \Big[a_i M_i^2 H^2 + b_i H^4 + c_i \frac{H^6}{M_i^2} + \dots
\Big].\end{equation} 
with the $\dots$ denoting higher (even) powers of $H$. Here, the Hubble parameter $H(t)$ plays the role of a ``running scale'' $\mu$ of the RGE. The (dimensionless) coefficients $a_i, b_i, c_i \dots$ receive contributions from loop corrections of fermion 
and boson matter fields with different
masses $M_i$.  The fact that there are only even powers of $H$ in the expansion on the right-hand side of 
(\ref{rge}) is due to general covariance~\cite{sola}.  The equation of state of the vacuum is that of a cosmological ``constant'' $p^{\rm RVM} = -\rho^{\rm RVM}$, despite the fact that the energy density depends on the cosmic time $t$ through its dependence on $H(t)$~\footnote{We note at this point that such an equation of state, corresponding to a time dependent vacuum energy, also characterises certain non-critical string/brane cosmologies~\cite{ncstring}.}.

The total energy density $\rho^{\rm total}$ of the cosmic fluid in the RVM framework, including radiation and matter (dust) contributions, is given by
\begin{equation}\label{rhorvm}
\rho^{\rm total} = \rho^\Lambda_{\rm RVM} + \rho^{\rm dust} + \rho^{\rm radiation}\,,
\end{equation}
where the vacuum contribution $\rho^\Lambda_{\rm RVM}$ is obtained by integrating (\ref{rge}):
\begin{equation}\label{rLRVM}
\rho^{\Lambda}_{\rm RVM}(H) \equiv \frac{\Lambda(H)}{\kappa^2}=
\frac{3}{\kappa^2}\left(c_0 + \nu H^{2} + \alpha
\frac{H^{4}}{H_{I}^{2}} + \dots \right) \;.
\end{equation}
In the above expression, the $\dots$ denote higher even powers of $H$, $\kappa^2 = 8\pi/M^2_{\rm P} = M_{\rm Pl}^{-2}$ is the four space-time dimensional gravitational constant, with $M_{\rm P}$ ($M_{\rm Pl} = 2.4 \times 10^{18}$~GeV) the (reduced) Planck mass, and $H_I$ is the Hubble Inflationary scale, taken to be in the ballpark of the Grand Unified Theories (GUT) scale, as suggested by the cosmological data~\cite{planck}; 
$c_0$ is an integration constant (with mass dimension $+2$ in
natural units), which can play the r\^ole of the current-epoch cosmological constant, and the coefficients 
$\nu$ and $\alpha$ can be viewed as the
reduced (dimensionless) beta-functions
of $\rho^{\Lambda}_{\rm RVM}$ at low and high energies respectively,
and are expected to be quite small in a typical GUT,
namely ${\cal O} (10^{-6}-10^{-3})$~\cite{sola}. In \eqref{rLRVM}, we truncated the $H$-expansion to $H^4$, since this suffices to describe the evolution of the Universe from inflation to the current era~\cite{history,bms}. In the early universe, the RVM energy-density term proportional to $H^4$ dominates and drives inflation~\cite{history}, but without an external inflaton field. The exit from this phase is smooth, but does not involve any reheating of the Universe, which in the standard cosmology would have 
occurred due to the inflaton decay. In this sense, the RVM evolution is distinct from that in standard cosmology.

The RVM framework has also been argued, not only to be consistent with the plethora of the cosmological data, but also to provide an alleviation of the $H_0$ and $\sigma_8$ tensions~\cite{tension}, and in this sense our string-inspired cosmological model, which is characterised by an effective energy density of RVM type~\cite{bms}, may be considered as providing an alternative to the standard concordance model that can also account for the observed tensions. 

The structure of the talk is as follows: in the next section \ref{sec:model}, I describe briefly the string-inspired cosmological model of \cite{bms} and its most important features; in section \ref{sec:infl}, I discuss how an inflationary phase is induced dynamically, through the formation of gravitational-anomaly condensates induced by primordial gravitational waves, while in the subsequent section \ref{sec:axion}, I describe briefly how leptogenesis occurs in this framework, in models with RHN, and how the primordial massless axions of the model acquire a mass, during the QCD era, which make them a potential candidate for DM. 
Some phenomenological features of the model are also discussed briefly. Finally, section \ref{sec:concl} contains our conclusions and outlook.
 
 \section{The String-Inspired Cosmological Model \label{sec:model}}
 
 In string theory~\cite{strings}, compactified to four dimensional space-times, the bosonic massless gravitational multiplet of the string spectrum consists of dilatons $\Phi$, gravitons $g _{\mu\nu}$ and antisymmetric tensor (Kalb-Ramond (KR)) spin-one fields $\mathcal B_{\mu\nu}$.
 The field strength of the latter, 
 ${\mathcal H}_{\mu\nu\rho} = \partial_\mu \mathcal B_{\nu\rho} + {\rm [cyclic~permutation~of~indices]}$,  is ``dual''~\cite{aben}  to a pseudoscalar field $b(x)$ (gravitational or KR axion): 
 \be\label{dualH}
 e^{-2\Phi} {\mathcal H}_{\mu\nu\rho} \propto \varepsilon_{\mu\nu\rho\sigma} \, \partial^\sigma b(x), 
 \ee
 with $\eps_{\mu\nu\rho\sigma}$ the covariant Levi-Cicita tensor density, with the Greek indices being world indices of the four-dimensional space-time manifold. It also satisfies the modified Bianchi identity~\cite{strings}:
\be\label{modbianchi2}
 \varepsilon_{abc}^{\;\;\;\;\;\;\;\mu}\, {\mathcal H}^{abc}_{\;\;\;\;\;\; \;\;\; ; \mu}
 =  \frac{\alpha^\prime}{32\, \kappa} \, c_1\, \sqrt{-g}\, \Big(R_{\mu\nu\rho\sigma}\, \widetilde R^{\mu\nu\rho\sigma} -
F_{\mu\nu}\, \widetilde F^{\mu\nu}\Big) 
\ee
where the tilde denotes the corresponding dual fields, 
the symbol $;$ denotes the gravitational covariant derivative, and Latin indices refer to the tangent-space of the world manifold at a point $x^\mu$.  The right-hand side of \eqref{modbianchi2} is the mixed gravitational and gauge anomaly, 
due to the addition of local counterterms, motivated by the requirement of anomaly cancellation~\cite{strings}.
 These are total derivative terms:
\begin{align}\label{pontryaginA}
&\sqrt{-g} \, c_1 \, \Big(R_{\mu\nu\rho\sigma}\, \widetilde R^{\mu\nu\rho\sigma} - F_{\mu\nu}\, \widetilde F^{\mu\nu} \Big) = \sqrt{-g} \, {\mathcal K}_{\rm mixed}^\mu (\omega)_{;\mu} = \partial_\mu \Big(\sqrt{-g} \, {\mathcal K}_{\rm mixed}^\mu (\omega) \Big)\nonumber \\ &
= 2 \, c_1 \,  \partial_\mu \Big[\epsilon^{\mu\nu\alpha\beta}\, \omega_\nu^{ab}\, \Big(\partial_\alpha \, \omega_{\beta ab} + \frac{2}{3}\, \omega_{\alpha a}^{\,\,\,\,\,\,\,c}\, \omega_{\beta cb}\Big)  - 2 \epsilon^{\mu\nu\alpha\beta}\, \Big(A^i_\nu\, \partial_\alpha A_\beta^i + \frac{2}{3} \, f^{ijk} \, A_\nu^i\, A_\alpha^j \, A_\beta^k \Big)\Big],
\end{align}
with Latin letters $i,j,k$ being gauge group indices; $\omega_\mu^{ab}$ is the spin-connection one-form and $A_\mu^i$ are gauge fields. The quantity $\sqrt{-g}\, {\mathcal K}_{\rm mixed}^\mu$ is the mixed-anomaly current density. 
The constant $c_1 = N_f$ in \eqref{modbianchi2} and \eqref{pontryaginA} denotes the number of chiral fermion species $N_f$ that contribute to the anomalous loops. Its precise value  depends on the underlying microscopic content of the theory. 

We now remark that, contrary to the traditional string approach, in ref. \cite{boss} we have not insisted on anomaly cancellation but have kept the (four-space-time dimensional) anomaly terms in \eqref{modbianchi2}.This constitutes the basis for our novel cosmology developed in ref.~\cite{bms}. In the latter work we have made the further important assumption that {\it only} gravitational degrees of freedom characterise the very early Universe phase, after the Big Bang. This has to be understood as implying  the existence of non-gravitational (matter and radiation) fields only as virtual ones, i.e. appearing only in internal lines of the Feynman graphs of the (quantum string effective) gravitational field theory describing the very early Universe dynamics. Virtual chiral fermion matter is therefore present, and its 
circulation in the anomaly loops leads to the non vanishing of the right-hand side of \eqref{modbianchi2}, with the gauge field terms being absent in the very early cosmic eras. As we shall discuss later on, in section \ref{sec:axion}, at the end of inflation chiral matter generation implies that fermion currents are also allowed to appear in external lines of graphs. This leads to the presence of additional terms in the effective action containing these chiral fermions, which have anomalous interactions with the KR axion $b(x)$. These 
can cancel the primordial gravitational anomalies during the radiation and matter eras~\cite{bms}, but gauge terms in the anomaly may now be present~\cite{bms}. For notational convenience, from now on we shall set $c_1=1$ by absorbing this coefficient in the pertinent anomaly currents, but we must always bear in mind that such a factor is actually present and depends on the details of the underlying microscopic model. 

In \cite{bms,boss} we implemented the constrain \eqref{modbianchi2} via a Lagrange multiplier field $b(x)$ in the respective path integral~\cite{kaloper}. Path-integration of the $\mathcal H_{\mu\nu\rho}$-terms in the partition function leads to the field $b(x)$ acquiring dynamics, which thus becomes the KR axion, that classically satisfies \eqref{dualH}. This procedure is called H-dualisation. To lowest non-trivial order in a derivative expansion (power series in the Regge slope $\alpha^\prime = M_s^{-2}$, where $M_s \ne M_{\rm P}$ is the string mass scale), the so-obtained ``H-dual'' low-energy (bosonic) string effective action  reads~\cite{bms,boss}
 \begin{align}\label{sea4}
S^{\rm eff}_B =&\; \int d^{4}x\sqrt{-g}\Big[ -\dfrac{1}{2\kappa^{2}}\, R + \frac{1}{2}\, \partial_\mu b \, \partial^\mu b
+   \sqrt{\frac{2}{3}}\,
\frac{\alpha^\prime}{96 \, \kappa} \, b(x) \, R_{\mu\nu\rho\sigma}\, \widetilde R^{\mu\nu\rho\sigma} + \dots \Big] \nonumber \\
=&\; \int d^{4}x\, \sqrt{-g}\Big[ -\dfrac{1}{2\kappa^{2}}\, R + \frac{1}{2}\, \partial_\mu b \, \partial^\mu b  -
 \sqrt{\frac{2}{3}}\,
\frac{\alpha^\prime}{96 \, \kappa} \, \partial_\mu b(x) \, {\mathcal K}^\mu + \dots \Big]~,
\end{align}
where in the second equality of \eqref{sea4} we have partially integrated the CP violating gravitational anomaly term, also termed 
gravitational Chern Simons term (\gCS), taking into account the total derivative nature \eqref{pontryaginA} of the anomaly. The  ${\mathcal K}^\mu$ is the \gCS current. The $\dots $ denote dilaton-derivative-dependent and higher derivative terms, which we shall not take into account in our discussion here. 
As discussed in \cite{bms}, we may consider constant dilatons throughout, in a self-consistent way.
On the other hand, independently of strings, one may consider the effective action \eqref{sea4} as the gravitational field-theory model, which our Cosmology will be based upon. 
As mentioned above, the gauge field anomaly terms are set to zero for the description of the early universe dynamics~\cite{bms}, but will play a r\^ole in the radiation era, to be discussed in section \ref{sec:axion}. 

It is important to note that in a Friedman-Lema$\hat i$tre-Robertson-Walker (FLRW) background space-time, of cosmological relevance,  the gravitational anomaly {\it vanishes}, but this is not the case in the presence of metric fluctuations, such as those associated with primordial gravitational waves (GW)~\cite{stephon,bms}. Unlike the gauge anomaly term, the gravitational anomaly \gCS does contribute to the stress tensor, since its variation with respect to the metric field is non trivial, but gives rise to the so-called Cotton tensor,  $\mathcal C^{\mu\nu}$~\cite{jackiw}: 
$\delta \Big[ \int d^4x \sqrt{-g} \, b \, R_{\mu\nu\rho\sigma}\, \widetilde R^{\mu\nu\rho\sigma} \Big] = 4 \int d^4x \sqrt{-g} \, {\mathcal C}^{\mu\nu}\, \delta g_{\mu\nu} = -
4 \int d^4x \sqrt{-g} \, {\mathcal C}_{\mu\nu}\, \delta g^{\mu\nu}$, with 
 \begin{align}\label{cotton}
\mathcal C^{\mu\nu} &=  -\frac{1}{2} \Big[v_\sigma \, \Big( \varepsilon^{\sigma\mu\alpha\beta} R^\nu_{\, \, \beta;\alpha}  +
\varepsilon^{\sigma\nu\alpha\beta} R^\mu_{\, \, \beta;\alpha}\Big)   + v_{\sigma\tau} \, \Big(\widetilde R^{\tau\mu\sigma\nu} +
\widetilde R^{\tau\nu\sigma\mu} \Big)\Big]   \nonumber \\ &= - \frac{1}{2} \Big[\Big(v_\sigma \, \widetilde R^{\lambda\mu\sigma\nu}\Big)_{;\lambda}  + \, (\mu \leftrightarrow \nu)\Big]\, , \qquad  v_{\sigma} \equiv \partial_\sigma b = b_{;\sigma}, \,\,v_{\sigma\tau} \equiv  v_{\tau; \sigma} = b_{;\tau;\sigma}.
\end{align}
From \eqref{cotton}, it follows that the Cotton tensor is traceless~\cite{jackiw}
\begin{align}\label{tracecot}
g_{\mu\nu}\, \mathcal C^{\mu\nu}= 0~.
\end{align}
The presence of a non-trivial Cotton tensor leads to modified Einstein equations~\cite{bms}:
\begin{align}\label{einsteincs}
R^{\mu\nu} - \frac{1}{2}\, g^{\mu\nu} \, R - \sqrt{\frac{2}{3}}\,
\frac{\alpha^\prime\, \kappa}{12} \,{\mathcal C}^{\mu\nu} = \kappa^2 \, T^{\mu\nu}_{b},
\end{align}
where
\begin{align}\label{tbstress}
T_b^{\mu\nu} = \partial^\mu b\, \partial^\nu b -  \frac{1}{2}\, g^{\mu\nu}\,  \Big(\partial^\alpha b \, \partial_\alpha b\Big)~,
\end{align}
is the stress tensor of the massless KR axion,   

From the form of the Cotton tensor \eqref{cotton}, and using Einstein equations \eqref{einsteincs},  one can derive the property~\cite{jackiw}:  
\begin{align}\label{csder}
\sqrt{\frac{2}{3}}\,\frac{\alpha^\prime\, \kappa}{12} \, {\mathcal C}^{\mu\nu}_{\,\,\,\,\,\,\,\,\,\,\,\,\,;\mu} = -\sqrt{\frac{2}{3}}\,\frac{\alpha^\prime\, \kappa}{12} \, \frac{1}{8} (\partial^\nu b) \, R^{\alpha\beta\gamma\delta} \, \widetilde R_{\alpha\beta\gamma\delta} = - \kappa^2 \, T^{\mu\nu}_{b\,\,\,\,\,\,\,\,\,;\mu}~,
\end{align}
which implies that, in the presence of {\it gravitational} anomalies, the diffeomorphism invariance, and hence the conservation of $T^{\mu\nu}_b$ is affected. 
Nonetheless, there is no fundamental issue, since, as follows from \eqref{csder}, there is a {\it conserved}  modified stress-energy tensor
\begin{align}\label{cons}
\kappa^2 \, {\widetilde T}_{b + g\mathcal C\mathcal S}^{\mu\nu} \equiv \sqrt{\frac{2}{3}}\,\frac{\alpha^\prime\, \kappa}{12} \mathcal C^{\mu\nu} + \kappa^2 T_b^{\mu\nu} \quad \Rightarrow \quad  {\widetilde T}_{b + g\mathcal C\mathcal S \,; \mu}^{\mu\nu} =0~,
\end{align}
and hence, the non-vanishing divergence of the Cotton tensor in anomalous backgrounds simply expresses a non-trivial 
energy exchange between the axion $b$-field and gravity.
 
The modified Einstein equations \eqref{einsteincs} affect the cosmological evolution in space times with, {\it e.g.}, primordial GW perturbations, which imply non-trivial gravitational anomalies~\cite{stephon}, and in fact, as we discuss below, the vacuum energy density resembles~\cite{bms} that of a RVM~\cite{sola}, leading self consistently to dynamical inflation without the need for external inflaton fields.

 \section{Inflation through Gravitational Anomaly Condensates \label{sec:infl}}
 
 In a de Sitter-type cosmological  background, characterised by an (approximately)  constant Hubble parameter $H \simeq$ const., the average $\langle \dots \rangle$ of the anomalous g$\mathcal C\mathcal S$ term, 
over {\it primordial} GW  fluctuations of the metric tensor, yields a non-trivial result~\cite{stephon}:
\begin{align}\label{rrt}
  \langle R_{\mu\nu\rho\sigma}\, \widetilde R^{\mu\nu\rho\sigma} \rangle  = \frac{16}{a^4} \, \kappa^2\int_0^\mu \frac{ 4\pi \, k^2\, d k}{(2\pi)^3} \, \frac{H^2}{2\, k^3} \, k^4 \, \Theta + {\rm O}(\Theta^3) , \qquad 
 \Theta {\equiv }\sqrt{\frac{2}{3}}\, \frac{\alpha^\prime \, \kappa}{12} \, H \,  {\dot {\overline b}} \, \ll 1,
  \end{align}
under the
slow-roll assumption for the (homogeneous and isotropic) KR axion  field $b(t)$,
 \begin{equation}\label{slowroll}
{\dot {\overline b}} \ll  H/\kappa, 
\end{equation}
where the overdot denotes derivative with respect to the cosmic time.
Because the integral in \eqref{rrt} is quartically divergent, we use an ultraviolet cutoff $\mu$ for the momentum (Fourier) scale $k$. The notation $\overline b(t)$ in \eqref{rrt} indicates a background solution of the equations of motion for the KR field. 
The smallness of $\Theta$ {(i.e. $|\Theta|\ll 1$)}, when combined with \eqref{slowroll}, implies the {\it sufficient} condition
\begin{align}\label{HMs}
{H^2/M^2_s \ll 12\sqrt{3/2}\ \ \Rightarrow\ \ \ H/M_s \ll 3.83\,.}
\end{align}
In agreement with the cosmological data~\cite{planck}, we take  for concreteness the inflationary Hubble parameter in the range:
\begin{equation}\label{Hinfl}
\frac{H}{\MPl}\sim  10^{-4} ~,
\end{equation}
which, on account of \eqref{rrt} and \eqref{Hinfl}, implies the {\it sufficient} condition
\begin{align}
\frac{M_{\rm Pl}}{M_s} \ll 3.83 \times (10^4 - 10^5).
\end{align}

The classical equations of motion of the KR axion field $b(x)$, stemming from (\ref{sea4}),
\begin{align}\label{krbeom}
\partial_{\alpha}\Big[\sqrt{-g}\Big(\partial^{\alpha}\bar{b}  -  \sqrt{\frac{2}{3}}\,
\frac{\alpha^\prime}{96 \, \kappa} \, {\mathcal K}^{\alpha}  \Big)\Big] = 0,
\end{align}
imply, as partial solutions, homegeneous and isotropic backgrounds~\cite{bms}:
\begin{align}\label{krbeom2}
\dot{\overline{b}}  =  \sqrt{\frac{2}{3}}\, \frac{\alpha^\prime}{96 \, \kappa} \, {\mathcal K}^{0}.
\end{align}
From \eqref{pontryaginA}, adapted appropriately to a homogeneous and isotropic situation, it is straightforward 
to arrive at the following evolution equation for the temporal component of the anomaly current density~\cite{bms}:
 \begin{align}\label{k01}
& \frac{d}{dt}  \Big(\sqrt{-g}\, {\mathcal K}^0 (t) \Big) = \langle \sqrt{-g} \, R_{\mu\nu\rho\sigma}\, \widetilde R^{\mu\nu\rho\sigma} \rangle
 \simeq \sqrt{-g} \, \langle R_{\mu\nu\rho\sigma}\, \widetilde R^{\mu\nu\rho\sigma} \rangle \simeq \sqrt{-g} \, \frac{1}{\pi^2} \Big(\frac{H}{M_{\rm Pl}}\Big)^2 \, \mu^4\, \Theta  \nonumber \\& \simeq
 \Big[5.86 \times 10^{-5} \,  \Big(\frac{H}{M_{\rm Pl}}\Big)^3 \, \Big(\frac{\mu}{M_s}\Big)^4 \,  M_{\rm Pl}\Big]\, \times \, \Big(\sqrt{-g} \, {\mathcal K}^0 (t(\eta))\Big),
 \end{align}
where we used \eqref{rrt} and \eqref{krbeom2}. For approximately constant $H$, Eq. \eqref{k01} can be integrated to give 
\begin{align}\label{k02}
{\mathcal K}^0 (t(\eta))  &\simeq {\mathcal K}^0_{\rm begin} (t=0) \, \exp\Big[  - 3H\, t(\eta)\, \mathcal A\Big]~, \nonumber \\
{\mathcal A} &\equiv  1  -  1.95 \,  \times 10^{-5} \,  \Big(\frac{H}{M_{\rm Pl}}\Big)^2 \, \Big(\frac{\mu}{M_{s}}\Big)^4,
\end{align}
where we have set the beginning of inflation at $t=0$  and its end at $t \to +\infty$. Then, 
one observes~\cite{bms} that the momentum cutoff on the graviton modes is of order 
\begin{align}\label{transpl}
\frac{\mu}{M_s} \simeq 15 \, \Big(\frac{\MPl}{H}\Big)^{1/2} \quad \stackrel{\eqref{Hinfl}}{\Rightarrow} \quad 
 \mu \sim 10^{3} \, M_{s}.
 \end{align}
 This provides, through \eqref{krbeom2}, a self-consistent and necessary condition for ${\dot b}$ to be approximately constant during inflation,  which in turn implies that the KR axion background remains {\it undiluted} at the end of the inflationary period:
 \begin{align}\label{lvint}
 \dot{\overline b} = \sqrt{\frac{2}{3}}\, \frac{\alpha^\prime}{96 \, \kappa} \, {\mathcal K}^{0} \simeq {\rm constant} \quad \Rightarrow \quad 
 {\overline b}(t) = {\overline b}(0) + \sqrt{\frac{2}{3}}\, \frac{\alpha^\prime}{96 \, \kappa} \, {\mathcal K}^{0}\, t~.
 \end{align}
 Above, $\overline b(0)$ is an initial value of the KR axion field, at the beginning of inflation, immediately after the Big Bang.
 In the low-energy field theory limit of strings, we are dealing with here and in \cite{bms}, this is a phenomenological parameter. 
 The reader should take notice of the fact that the solution \eqref{lvint}  violates (spontaneously) Lorentz symmetry, and its existence is linked to the presence of a GW-induced {\it anomaly condensate} \eqref{rrt}.

In view of \eqref{HMs} and \eqref{Hinfl}, one obtains from \eqref{transpl} the sufficient condition
 \begin {align}\label{murange}
 \mu \gg 2.61 \times (10^{-3} - 10^{-2}) \, M_{\rm Pl},
 \end{align}
implying that the cutoff scale $\mu$ can be at least of order of $M_{\rm Pl}$. If one makes the reasonable assumption that in a consistent low-energy theory of quantum gravity transplanckian modes should be avoided, then, from \eqref{murange}, we obtain   
the following range for  the string scale~\cite{bms}
 \begin{align}\label{msr}
 \MPl \, \gtrsim  \, M_s \gtrsim 10^{-3} \, M_{\rm Pl}~,
 \end{align}
which guarantees the existence of the Lorentz-violating solution \eqref{lvint} for the KR background.

We now remark that a slow-roll condition on the KR background, as dictated  by the Planck data~\cite{planck}, is consistent with \eqref{lvint}~\cite{bms}:
\begin{align}\label{slowrollepsi}
 \epsilon \sim \frac{1}{2} \frac{1}{(H M_{\rm Pl})^2}\, {\dot {\overline b}}^2 \sim
 10^{-2} \quad \Rightarrow \quad {\overline b}(t) \, \sim \, \overline b(0) + \sqrt{2\,\epsilon} \, M_{\rm Pl} \, H\, t\,.
\end{align}
From \eqref{slowrollepsi}, \eqref{Hinfl}, then, we obtain for the 
anomaly condensate ${\mathcal K}^0(0)$:
\begin{align}\label{k0}
{\mathcal K}^0(0) \sim  0.00134 \, M_s^2 \, M_{\rm Pl}
\end{align}
which, on account of \eqref{msr} lies in the range 
\begin{align}\label{k0range}
1.34 \cdot 10^{-3}  \gtrsim \frac{{\mathcal K}^0(0)}{\MPl^3}  \gtrsim  1.34 \cdot 10^{-9}.
\end{align}

So far, we have assumed the existence of a de Sitter background, whose origin was left unspecified. In our context, such a space-time is induced {\it dynamically} as a consequence of the GW condensate~\cite{bms}. Indeed, the presence of GW fluctuations leads to the following non-vanishing result for the averaged quantity 
\begin{align}\label{order}
\langle g\mathcal C\mathcal S \rangle & = \sqrt{\frac{2}{3}}\, \frac{\alpha^\prime}{96 \, \kappa} \, \int d^4x \, \sqrt{-g}\, \langle \, \overline b(x) \,  R_{\mu\nu\rho\sigma}\, \widetilde R^{\mu\nu\rho\sigma} \, \rangle
\nonumber \\ & \simeq   \int d^4 x \, \sqrt{-g}\, \Big(5.86 \times 10^7 \, \, \sqrt{2\, \epsilon} \,
\Big[\frac{\overline b(0)}{\MPl} + \sqrt{2\, \epsilon} \,  \mathcal N \Big] \, H^4 \Big) \,
\equiv  - \int d^4x \, \sqrt{-g} \, {\rho_{\Lambda\,cond}}~,
\end{align}
where $\langle \dots \rangle$ denotes average over GW graviton fluctuations about a de Sitter background. In the second line we have used \eqref{slowrollepsi}, and a maximum order of magnitude esrtimate~\cite{bms} for $\overline b(t)$ 
evaluated at the end of the inflationary period, $t_{\rm end}$, for which $H\, t_{\rm end} \sim {\mathcal N}$ 
with ${\mathcal N}$ the number of e-foldings; phenomenologically~\cite{planck}, $\mathcal N$ may be taken in the range  
$\mathcal N \in [60-70] $. It is then estimated that to have an approximate constant (de Sitter) type positive cosmological constant term, one needs (in our conventions): $\overline b(0) < 0$, with 
\begin{align}\label{b0}
|\overline b(0)| \gtrsim \sqrt{2\, \epsilon}\, {\mathcal N} \, \MPl~.
\end{align}
For our purposes it suffices to take 
\begin{align}\label{b02}
|\overline b(0)|\sim 10 \, \MPl.
\end{align}

The term \eqref{order}, then, contributes a positive cosmological constant type term in the effective action, given that the variation of the anomaly condensate $\langle \, \overline b(x) \,  R_{\mu\nu\rho\sigma}\, \widetilde R^{\mu\nu\rho\sigma} \, \rangle$ with respect to the metric tensor vanishes, as happens with all condensates. In a sense, the appearance of a condensate, which is a purely quantum phenomenon, breaks the scale symmetry of the model \eqref{sea4}, and introduces a non-vanishing ``trace'' for the Cotton tensor, which classically is traceless \eqref{tracecot}.  The reader should notice the non-linear $H^4$ dependence of this term.

From the (free) stress tensor of the massless KR axion we also obtain
$\rho^{b} = \frac{1}{2} (\dot {\overline b})^2 \simeq  \epsilon \, M_{\rm Pl}^2 \, H^2$
with a ``stiff-matter'' equation of state $w_b=1$.  Using appropriately \eqref{pontryaginA} and \eqref{csder}, for a homogeneous and isotropic situation, {\it i.e.}
\begin{align}\label{solution}
C^{\mu 0}_{\,\,\,\,\,\,\,\,\,\,;\mu} &= \frac{d}{dt} \mathcal C^{00} + 4 H\, \mathcal C^{00}  \simeq - \frac{1}{8} \, \dot{\overline b} \, \langle R^{\alpha\beta\gamma\delta} \, \widetilde R_{\alpha\beta\gamma\delta}\rangle
 \simeq  \, - \frac{1}{8} \sqrt{\frac{2}{3}}\, \frac{\alpha^\prime \, \kappa}{12} \, H \,
\frac{1}{\pi^2} \Big(\frac{H}{M_{\rm Pl}}\Big)^2 \, \mu^4 \, {\dot {\overline b}}^2,
 \end{align}
it is straightforward 
to arrive at the following evolution equation for the energy densities of the various components $\rho^b$, 
$\rho^{g\mathcal C \mathcal S} \equiv \mathcal C^{00}$ and $\rho_\Lambda$, of the string universe during the early GW condensate phase~\cite{bms}
\begin{align}\label{cons3}
\frac{d}{dt}(\rho^b + \rho^{g\mathcal C\mathcal S}) + 3 H \Big( (1+w_b)\, \rho^b + \frac{4}{3}\rho^{g\mathcal C\mathcal S} \Big)  \simeq 0 \quad  \Rightarrow \quad
\rho^b \simeq -\frac{2}{3}\rho^{g\mathcal C\mathcal S}~, 
\end{align}
where the last result holds, because, during the de Sitter phase one has approximately $\frac{d}{dt}(\rho^b + \rho^{g\mathcal C\mathcal S}) \simeq 0$. 
The total energy density
\begin{align}\label{toten}
\rho_{\rm total} &= \rho_b + \rho_{g\mathcal C\mathcal S} + \rho_{\Lambda \, cond} \nonumber \\& \simeq
3\MPl^4 \, \Big[ -1.7 \times 10^{-3} \Big(\frac{H}{\MPl}\Big)^2
+ \frac{\sqrt{2}}{3} \, \frac{|\overline b(0)|}{\MPl}\, \times {5.86\, \times} \, 10^6 \, \left(\frac{H}{\MPl}\right)^4 \Big] > 0~.
\end{align}
is {\it positive} and drives the de Sitter (inflationary) space-time due to the $H^4$ term, without the need for external inflatons. In fact, the form of the total energy density resembles that of an RVM \eqref{rLRVM}, however the coefficient $\nu$ is negative in our case, during the inflationary phase. The coefficient $\alpha$ on the other hand is positive and this term is dominant in the early Universe and drives inflation, in agreement with generic properties of RVM~\cite{history}. In view of \eqref{b02}, we estimate
$\alpha= \frac{\sqrt{2}}{3} \, \frac{|\overline b(0)|}{\MPl} \, \times \, {5.86\, \times} \,10^6 \left (\frac{H_I}{\MPl}\right)^2\sim 2.8 \times 10^{-1}$, for an inflationary scale Hubble, $H_I$ near a GUT scale~\cite{planck}. In fact, such values are consistent with generic predictions from RVM~\cite{sola}; indeed, if one identifies the total energy density during inflation, which is dominated by $\rho_{\Lambda \, cond}$,  with a GUT potential,  $V \sim M_X^4$, corresponding to an energy scale $M_X$,
then~\cite{bms}
\begin{align}\label{mxscale}
\rho_{\rm total} 
\simeq {\frac{{|\overline b(0)|}}{\MPl}\, 8.3\times10^{-10}\, \MPl^4}
\,  \Rightarrow \, M_X {\simeq 1.3 \, \times 10^{16}\,  \Big(\frac{{|\overline b(0)|}}{\MPl} \Big)^{1/4}~ {\rm GeV} \simeq 2.3 \, \times 10^{16}~ {\rm GeV}\,,}
\end{align}
using \eqref{b02}, which is thus in agreement with RVM phenomenology based on GUTs.

The evolution of this string-inspired Universe does not follow the standard smooth RVM evolution, although the structure of its vacuum energy density falls into this category. The coefficients $\nu$ and $\alpha$ are different for the various epochs, since new contributions from various matter and radiation sources arise after the exit from inflation. As discussed in some detail in \cite{bms}, for the current era, the corresponding coefficient $\nu_0$ of the $H_0^2$ terms in the vacuum energy density becomes positive, receiving contributions from traditional cosmic electromagnetic fields. 

The generation of an axion potential,  say by non perturbative effects in the matter era, as we shall discuss in the next section, helps in ensuring the validity of a (slowly moving) LV background configuration for the $b$ field in the current epoch, of the form \eqref{lvint},
\begin{align}\label{today}
\dot{\overline b}|_{\rm today} \simeq \sqrt{2\,\epsilon^\prime} \, H_0 \, \MPl
\end{align}
where the suffix $0$ indicates present-day quantities, and $\epsilon^\prime$ is a slow-roll parameter for the current era, which is in general different from $\epsilon$ appearing in \eqref{slowrollepsi} during the inflationary phase, although current phenomenology dictates that one should expect~\cite{bms}
\begin{align}\label{epsprime}
\epsilon^\prime \sim \epsilon = \mathcal O(10^{-2}). 
\end{align}
since in this case one could obtain a present-epoch DM content in the phenomenologically right ball park~\cite{planck}
\begin{equation}\label{eq:UbKb}
  \Omega_{m0}=\frac{\rho_{m0}}{\rho^{(0)}_c}\simeq \frac{U_b}{\rho^{(0)}_c}\simeq 10\, \frac{K_b}{\rho^{(0)}_c}\simeq 10\, \epsilon^\prime={\cal O}(0.1)\,.
\end{equation}
Above, $\rho_{m0}$ denotes the current energy density of DM in the universe, and we used the fact that
the slow-roll parameter $\epsilon^\prime$ in \eqref{today} measures the ratio of the kinetic energy of the axion field $b(x)$,  
$K_b\sim (1/2)\,\dot{b}^2$, to the current critical energy density of the Universe, $\rho^{(0)}_c=(M_{\rm Pl} H_0)^2/3$. Following expectations from generic quintessence models, we assumed in \eqref{eq:UbKb} that the the potential $U_b$ of the axion during the present era is an order of magnitude larger than its kinetic energy $K_b$. Precise justification of \eqref{epsprime}, based on microscopic considerations, is still pending. In \cite{bms}, we presented a scenario in which the dominant contribution to $\epsilon^\prime$ comes from a present-era cosmic magnetic field, but the intensity of such a field 
is, at present, treated as a phenomenological parameter, awaiting precise determination from underlying detailed string theory models, which is not a trivial task.

Despite the aforementioned evolutionary differences of our model~\cite{bms} from the traditional RVM, there are important similarities. Indeed, as in the traditional RVM~\cite{sola,history}, the inflationary phase in our string-inspired model is induced by the dominant $H^4$ term in \eqref{toten}, due to the GW-anomaly condensate, without the need for an external inflaton field, and there is no traditional reheating of the Universe during the exit phase. However, there is a novel feature, which is specific to our model~\cite{bms}, not characterising the generic RVM. In our case, as discussed above, there is an undiluted LV axion background which carries over to the radiation phase, playing a non-trivial r\^ole in generating matter-antimatter asymmetry and axion DM, as we next proceed to discuss. 

 \section{Radiation-Matter Eras: Matter-Antimatter Asymmetry and Axion Dark Matter \label{sec:axion}}

At the end of inflationary era (exit phase) of our string-inspired Universe, ordinary chiral fermionic matter is generated~\cite{bms}, which is anomalous in the sense of being chraracterised by a {\it non-conserved} axial fermion matter current $J^{5\,\mu} = \sum_i \overline \psi_i^{\rm ch} \gamma^\mu\, \gamma^5 \, \psi^{\rm ch}_i $, where $\psi^{\rm ch}_i$ denote species of chiral fermions. The effective string action  in such an era reads:
\begin{align}\label{sea6}
S^{\rm eff} &=\; \int d^{4}x\sqrt{-g}\Big[ -\dfrac{1}{2\kappa^{2}}\, R + \frac{1}{2}\, \partial_\mu b \, \partial^\mu b   -  \sqrt{\frac{2}{3}}\,
\frac{\alpha^\prime}{96\, \kappa} \, \partial_\mu b(x) \, {\mathcal K}^\mu
\Big]  \nonumber \\
& + S_{fermion}^{Free} + \int d^{4}x\sqrt{-g}\,  \frac{\alpha^\prime}{\kappa} \, \sqrt{\frac{3}{8}} \, \partial_{\mu}b \, J^{5\,\mu}    - \dfrac{3\, {\alpha^\prime}^{2}}{16\, \kappa^2}\, \int d^{4}x\sqrt{-g}\,J^{5}_{\,\,\mu}\,J^{5\,\mu}  + \dots,
\end{align}
where the four fermion-axial-current-current term is an indication of the existence of ``torsion'' in the geometry, due to the KR field strength, for details see refs.~\cite{bms,boss}. 

The basic assumption of our scenario is the {\it cancellation} of the gravitational anomalies that existed in the inflationary effective action by those generated due to the chiral matter, in the sense that~\cite{bms}
\begin{align}\label{anom2}
& \partial_\mu \Big[\sqrt{-g}\, \Big(  \sqrt{\frac{3}{8}} \frac{\alpha^\prime}{\kappa}\, J^{5\,\mu}  -  \frac{\alpha^\prime}{\kappa}\, \sqrt{\frac{2}{3}}\,
\frac{1}{96} \, {\mathcal K}^\mu  \Big) \Big]   =   \sqrt{\frac{3}{8}} \, \frac{\alpha^\prime}{\kappa}\, \Big(\frac{\alpha_{\rm EM}}{2\pi}  \, \sqrt{-g}\,  {F}^{\mu\nu}\,  \widetilde{F}_{\mu\nu} + \frac{\alpha_s}{8\pi}\, \sqrt{-g} \, G_{\mu\nu}^a \, \widetilde G^{a\mu\nu} \Big)~,
\end{align}
where $F_{\mu\nu}$ is the electromagnetic (EM) Maxwell tensor,  associated with electromagnetic cosmic fields and $G_{\mu\nu}^a$ is the gluon field strength of QCD, with $a=1, \dots 8$ an adjoint SU(3) colour index; the tilde denotes the corresponding dual tensors;  $\alpha_{\rm EM}$ is the electromagnetic fine structure constant, and $\alpha_s$ is the strong interactions fine structure constant, and their presence is associated with the fact that the anomalies are one-loop (exact) effects. In the radiation and matter eras gauge fields are present, and hence we use the mixed anomaly terms ({\it cf.} \eqref{pontryaginA}) in the non-trivial divergence of the axial chiral-fermion matter current $J^{5\, \mu}$.~\footnote{The reader should notice that the number of chiral fermion species $N_f$ contributing to the triangle anomalies, which depends on details of the undelying microscopic model, is not shown explicitly in \eqref{anom2}, as it has been absorbed (as a common proportionality factor) in the expression for the gravitational anomaly current $\mathcal K^\mu$, as well as in the gauge-coupling structure constants on the right-hand side ({\it cf.} discussion below \eqref{modbianchi2} in section \ref{sec:intro}).}

During the radiation era, the undiluted KR axion LV background \eqref{lvint} can induce tree-level leptogenesis in models with {\it massive} right-handed neutrinos (RHN) according to the mechanism suggested in \cite{boss}, and presented by the speaker in the {\it 2018 edition of the Corfu Schools and Workshops}~\cite{corfu18}. Decays of such RHN into standard model particles and antiparticles due to Higgs portal interactions, in the presence of such backgrounds \eqref{lvint}, induce lepton asymmetries proportional to the (approximately constant) rate of change of the background, $\dot{\overline b}$. The lepton asymmetries are then communicated to the baryon sector via Baryon-minus-Lepton-number conserving sphaleron processes in the SM sector~\cite{kuzmin}, resulting in matter-antimmater asymmetry.
The observed value of the asymmetry~\cite{planck} can be obtained by fixing the value of the background. In \cite{bms} 
it was shown that, remarkably, upon taking into account the mild temperature dependence of the KR background due to thermal effects of the expanding Universe in the radiation era~\cite{boss}, the solution \eqref{lvint}, with the restriction \eqref{msr},  yields phenomenologically consistent results for the scenario of \cite{bms}, in which leptogenesis occurs at decoupling temperatures $T_D \sim 10^5$~GeV and RHN masses $m_N \sim T_D$, which are also the conditions of \cite{boss}. 

During the QCD era, that is the epoch of the expanding Universe in which the temperature is of the order of a few hundreds of MeV (order of the QCD scale), it is possible that non-perturbative instanton effects generate a potential for the KR axion $b(x)$
\begin{align}\label{vqcd}
V_b^{\rm QCD}  \simeq \Lambda^4_{\rm QCD} \Big(1 - {\rm cos}(\frac{b}{f_b})\Big)~,  \quad 
 f_b &\equiv \sqrt{\frac{8}{3}} \, \frac{\kappa}{\alpha^\prime} =  \sqrt{\frac{8}{3}} \, \Big(\frac{M_s}{\MPl}\Big)^2\, \MPl ~.
\end{align}
where $\Lambda_{\rm QCD} \sim 218~{\rm MeV}$ is the QCD scale, and $f_b$ plays the r\^ole of the axion decay constant. For the range of the string scale \eqref{msr}, which guarantees the LV solution \eqref{lvint} for the KR background, the resulting range for the axion decay constant $f_b$ in \eqref{vqcd} is~\cite{bms} 
\begin{align}\label{fbr}
3.9 \times 10^{12} ~{\rm GeV}\lesssim\, f_b \,\lesssim\,3.9 \times 10^{18} ~{\rm GeV}\,.
\end{align}
Phenomenologically, the ordinary QCD axion decay constant $f_a$ is found to lie in the range 
 $10^9~{\rm GeV} < f_a < 10^{12}~{\rm GeV}$, and thus from \eqref{fbr}, we observe  that there is a marginal overlap. If the constraints of \cite{astro}, however, are taken into account, the overlap of the allowed regions between $f_a$ and $f_b$ increases significantly.
 
The potential \eqref{vqcd} also implies the generation of KR-axion mass during the QCD epoch, 
\begin{align}\label{axionmass}
m_b &=  \frac{\Lambda^2_{\rm QCD}}{f_b} = \sqrt{\frac{3}{8}} \, \Big(\frac{\Lambda_{\rm QCD}}{M_s}\Big)^2\,  \MPl
= \sqrt{\frac{3}{8}} \, \Big(\frac{\Lambda_{\rm QCD}}{\MPl}\Big)^2 \, \Big(\frac{\MPl}{M_s}\Big)^2\,  \MPl ~,
\end{align}
which, in view of \eqref{msr}, lies in the range
\begin{align}
{ 1.17 \times 10^{-11} ~{\rm  eV} \lesssim m_b \, \lesssim \, 1.17 \times 10^{-5}~{\rm  eV} \,,}
\end{align}
in agreement with the order of magnitude of axion masses calculated in lattice QCD approaches~\cite{latticeqcd}: 
$m^{\rm lattice}_a \sim 5.7 \, (\frac{10^{12}~{\rm GeV}}{f_a}) \, \times 10^{-6}$~eV. Such axions may play the r\^ole of a DM component.  

 In ref.~\cite{bms} we also discussed more complicated scenarios for generating KR axion masses, in agreement with  KR
 background LV solutions of the form \eqref{lvint}, which involve mixing of the KR axion with other axions that are abundant in string theories, stemming from appropriate string flux fields~\cite{arvanitaki}. The mixing could be, {\it e.g.}, of a kinetic form discussed in \cite{mp}, which, by the way, provides a scenario for the generation of anomalous radiative masses for the RHN due to the quantum fluctuations of the KR axion itself. Such extended models also allow for ultralight (in our case, stringy in origin) axion DM, which is currently being searched actively with diverse methods, some of them employing innovative quantum-sensor technologies~\cite{sensors}.

\section{Conclusions and Outlook \label{sec:concl}} 

In this work we have discussed recent developments in a scenario~\cite{bms} for a cosmological evolution of a string-inspired model for the Universe, in which gravitational anomalies, induced by primordial gravitational waves in the very early Universe, can induce inflation dynamically and exit from it, in a way predicted in the framework of running vacuum models (RVM). 

Although there are some distinct features between our model and a generic RVM, concerning, in particular, the form and order of magnitude of the  coefficients of the various even powers of the Hubble parameter in the expression for the vacuum energy density of the model, which differ from era to era, nonetheless the essential features of the RVM framework are there. Namely, there exists dynamical inflation without an external inflaton field, due to the presence of the higher powers of the Hubble parameter in the energy density, and exit without the traditional reheating, as well as a de Sitter-like equation of state for the vacuum contributions. 

There are also some novel important features specific to our model, like the existence of Lorentz-violating primordial axion backgrounds, from the gravitational massless (bosonic) multiplet of the string; these carry over undiluted onto the radiation phase, and play an important r\^ole in generating a matter-antimatter asymmetry in a novel way, in models with heavy right-handed neutrinos. Moreover, such axions can source Dark Matter,  provided appropriate potentials and masses are generated via non-perturbative effects during the radiation and matter eras. 

The cosmological model we presented has placed gravitational anomalies at a very central stage.
Such anomalies, induced by gravitational-wave metric fluctuations at the very early Universe,  
are essential for inducing inflation and the (undiluted) Kalb-Ramond (stringy in origin)  axion backgrounds. Although such anomalies are cancelled at the end of inflation, by the generation of chiral matter, nevertheless the fact that the undiluted Kalb-Ramond axion background, which has been argued above to be important for the generation of matter-antimatter asymmetry and dark matter in the Universe, owes its existence to them, leads us to conclude that our current existence might well be due to such anomalous interactions during the birth and infancy of our Universe. So paraphrasing Carl Sagan~\cite{sagan} the upshot of our work is that {\it ``we might well be anomalously made of starstuff''}~\cite{bms}. 
 
There are several avenues for further research in this model that we would like to pursue. First of all, we need to understand better the current era of the Universe, and justify in a microscopic way Eq.~\eqref{epsprime}. Moreover, since in string theory, dilatons play an essential r\^ole, we would like to study in general which kind of dilaton potentials of effective (supergravity) field theories, stemming from strings, are compatible with our scenario, which so far relied on constant dilatons. The introduction of non-trivial, time dependent dilatons at an early stage of the Universe evolution, that could characterise more general models, should be studied in detail. Such a question is non trivial but worths the effort, as it might lead to interesting early Universe effects that could be testable.
Last, but not least, a detailed phenomenology of the model against the plethora of cosmological and  astrophysical data, including issues like the alleviation of tensions in cosmological data, or inclusion of self interactions of DM, which our axion DM models do have, and which - as we have discussed above - might play a r\^ole in galactic structure, are among our immediate priorities.  We hope to be able to report on progress made in some of these issues soon.

\end{document}